# Unsupervised Discovery of Recurring Speech Patterns Using Probabilistic Adaptive Metrics

*Okko Räsänen*[1,2] *& María Andrea Cruz Blandón*[1]

[1]Unit of Computing Sciences, Tampere University, Finland
[2]Dept. Signal Processing and Acoustics, Aalto University, Finland
okko.rasanen@tuni.fi, maria.cruzblandon@tuni.fi

## Abstract

Unsupervised spoken term discovery (UTD) aims at finding recurring segments of speech from a corpus of acoustic speech data. One potential approach to this problem is to use dynamic time warping (DTW) to find well-aligning patterns from the speech data. However, automatic selection of initial candidate segments for the DTW-alignment and detection of "sufficiently good" alignments among those require some type of pre-defined criteria, often operationalized as threshold parameters for pair-wise distance metrics between signal representations. In the existing UTD systems, the optimal hyperparameters may differ across datasets, limiting their applicability to new corpora and truly low-resource scenarios. In this paper, we propose a novel probabilistic approach to DTW-based UTD named as PDTW. In PDTW, distributional characteristics of the processed corpus are utilized for adaptive evaluation of alignment quality, thereby enabling systematic discovery of pattern pairs that have similarity what would be expected by coincidence. We test PDTW on Zero Resource Speech Challenge 2017 datasets as a part of 2020 implementation of the challenge. The results show that the system performs consistently on all five tested languages using fixed hyperparameters, clearly outperforming the earlier DTW-based system in terms of coverage of the detected patterns.

## 1. Introduction

Unsupervised learning of language patterns and structures from acoustic speech data is a challenging problem. Systems capable of such processing could be utilized in low-resource speech processing scenarios where access to labeled data is limited (see, e.g., [1,2]). In addition, these systems can be used as models or learnability proofs for infant language learning, since infants also have to learn their native language without explicit supervision [3, 4]. In particular, the so-called Zero Resource Speech Processing initiative (www.zerospeech.com) has sought to drive the development and dissemination of new algorithms capable of unsupervised speech learning through a series of Zero Resource Speech Challenges (ZSC) [5–8]. While ZSC have evolved across the years, two main challenge tracks have persisted throughout all its implementations: Track-1 task focusing on automatic learning of phoneme-sensitive speech signal representations and Track-2 focusing on unsupervised spoken term discovery (UTD) from acoustic data.

In this paper, we describe our Speech and Cognition group (SPECOG) system for the 2020 implementation of the Zero Resource Challenge [8] that revisits the UTD task and associated datasets from the 2017 version of the challenge [6] with slightly revised evaluation protocols. More specifically, we revisit an old and widely used technique of dynamic time warping (DTW) [9] for finding recurring patterns in speech and propose a new probabilistic formulation for a DTW-based pipeline called as probabilistic DTW (PDTW). We show that PDTW enables consistent UTD performance across different corpora without adjusting the parameters of the system.

### 1.1. Earlier work on unsupervised speech pattern discovery

The main goal of UTD is to find pairs (or clusters) of speech segments that match in their phonemic content, be they syllables, words, or longer phrases. The primary problem is the enormous acoustic variability of speech signals, as the same spoken word never occurs twice in exactly the same acoustic form even when spoken by the same speaker. Different voices, speaking styles, background noises, recording setups, and many other factors all impose some type of changes to the signal characteristics in time and frequency. In addition, units such as words are not highlighted by any language-universal cues that would allow their accurate segmentation before the matching process. Together these factors make naïve approaches such as template or string matching techniques (e.g., [10–12]) applied on vector quantized data impractical for UTD. Instead, the potential solution must handle signal variation in time and frequency while deciding what counts as the "same" in speech.

The existing methods for unsupervised UTD can be broadly categorized into DTW-based and acoustic word embedding (AWE) -based approaches (but see also, e.g., [13]). The DTW-based approaches trace back to Segmental DTW algorithm by Park & Glass [14], followed by SWD-model by ten Bosch & Cranen [15], DP-ngram model by Aimetti [16], JHU-UTD system by Jansen and Van Durme [17], MODIS by Cantanese et al. [18] (see also [19]), and system by Räsänen and Seshadri [20] for the ZSC2017, to name a few. In theory, DTW is very powerful, since it enables pairwise alignment of any arbitrary multivariate patterns that may consist of non-linear temporal distortions with respect to each other, providing also a measure of the resulting alignment quality. However, exhaustive pairwise DTW between all possible segments of a corpus of moderate size is computationally infeasible, and practically all the proposed UTD methods use some kind of heuristics to limit the number DTW alignments carried out at full signal resolution. In addition, several hyperparameters are typically required to guide the alignment process and to ultimately select only those patterns that are sufficiently similar to each other according to the set criteria. The challenge in setting these parameters correctly becomes reflected in inconsistent performance of the DTW-methods across different corpora when using the same set of hyperparameters (see, e.g., [6]; see also Table 1).

In contrast to DTW-approaches, AWE-based systems aim to map variable-length speech patterns into a fixed-dimensional vector embedding space in which segment distance measurements or clustering operations can then be carried out using standard methods. Examples of AWE-based systems include a syllable-based n-gram system in [21] and ES-Kmeans algorithm by Kamper et al. [22]. The latter method can be considered as the current state-of-the-art in UTD, and it performs full segmentation of a corpus by simultaneously learning a segmentation and clustering solution that minimizes the joint cost of both operations. As a drawback of the full segmentation target, the parameters of the ES-Kmeans may be difficult to tune so that the discovered clusters are pure in terms of their phonemic content—a property that might be desirable for some applications.

In the present paper, we revisit the DTW-based approach for UTD and tackle the issue of hyperparameter setting across different speech corpora with our PDTW approach. We propose a two-stage DTW-based UTD approach that converts heuristic thresholds into a probabilistic interpretation: matching patterns are obtained by finding alignments that are highly unlikely to occur by chance, and where chance-level is defined in terms of the distributional characteristics of the given corpus.

## 2. Pattern discovery with PDTW

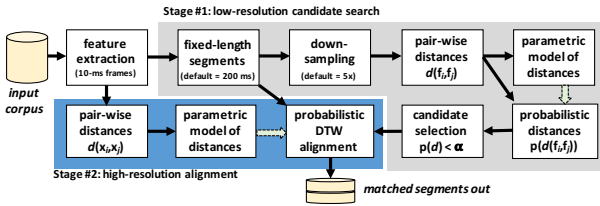

Figure 1: *A schematic view of the PDTW alignment algorithm.*

The overall structure of the PDTW pipeline is shown in Fig. 1, consisting of two main stages: In the first stage, an initial low-resolution matching of recurring segments is carried out to filter out a small number of promising alignment candidates. In the second stage, a high-resolution DTW is used to align all pair candidates from stage 1. The key idea is to replace heuristically defined similarity thresholds in both stages with probabilistic interpretations of pattern similarity: instead of thresholding frame-wise distances or alignment paths in some metric space, the distances are first converted into probabilities of observing such distances (or alignment paths) in the given dataset. Pattern selection is then carried out by selecting only those pattern pairs that are unlikely to have a given similarity measure by chance, where this probability of chance is a hyperparameter of PDTW.

### 2.1. Stage #1: low-resolution candidate search

First, input feature sequence $\mathbf{X} = [\mathbf{x}_1, \mathbf{x}_2, \ldots, \mathbf{x}_N]$, $\mathbf{x}_n \in \mathbb{R}^d$, corresponding to the entire speech corpus with $N$ frames is windowed into fixed-length segments of $L$ frames with a window shift of $S$ frames. Features of each segment are then uniformly downsampled to $M$ frames and concatenated, resulting in a new fixed-dimensional ($Md$ x 1) feature vector $\mathbf{f}_i$ for each segment $i$ (see also [21,22]). Cosine distances $d(\mathbf{f}_i, \mathbf{f}_j)$ between a number[1] of randomly sampled feature vector pairs are then measured, and a normal cumulative distribution $p(d_{i,j}) = \mathcal{N}_{\text{cdf}}(d(\mathbf{f}_i, \mathbf{f}_j) | \mu_f, \sigma_f^2)$ is then fit to the obtained distances. Finally, cosine distances from each vector $\mathbf{f}_i$ to all other vectors $\mathbf{f}_j$ are calculated, excluding the segments from temporal neighborhood of the current segment with $i \pm \{1, 2 \ldots N\} \neq j$, and information regarding up to $k$ nearest segments with $p(d_{i,j}) < \alpha$ are maintained in memory. The threshold parameter $\alpha$ can here be interpreted as the maximum allowed probability that the given distance is not smaller than the typical distances observed in the corpus, not unlike to significance criterion in a one-tailed statistical test. The underlying assumption is that if two segments at least partially share the same phonetic content, their mutual distance should be significantly lower than what is expected from random variation in the pairwise distances.

### 2.2. Stage #2: high-resolution alignment

In stage #2, candidate pairs from the first stage are analyzed more closely with DTW[2]. First, the original candidate segments are expanded by $\pm E$ frames to enable alignment of patterns up to a maximum duration of $L+2E$ frames. Then a regular affinity matrix $\mathbf{M}_{i,j}$ of the corresponding original segment features $\mathbf{X}_i$ and $\mathbf{X}_j$ is calculated using cosine-distance for each candidate pair. Since the stage #1 matching already requires that the segments are relatively well aligned, the best alignment path must lie close to the diagonal of $\mathbf{M}_{i,j}$. Now, instead of finding the shortest path through $\mathbf{M}_{i,j}$ using DTW, the pairwise cosine distances $d(\mathbf{x}_y, \mathbf{x}_z)$ (the elements of $\mathbf{M}_{i,j}$) are again (non-linearly) transformed into probabilities $p(d(\mathbf{x}_y, \mathbf{x}_z))$ that the distances are equal or larger to the typical distances in the corpus. Probabilities are obtained using a cumulative probability density function of normal distribution $\mathcal{N}_{\text{cdf}}(d(\mathbf{x}_y, \mathbf{x}_z) | \mu_d, \sigma_d^2)$ with mean $\mu_d$ and variance $\sigma_d^2$ estimated from a random sample[1] of paired feature vectors $\mathbf{x}_y$ and $\mathbf{x}_z$. DTW is then used to find a minimum cost path through the resulting probability matrix $\mathbf{P}_{i,j}$. Probability $p_{\text{align}}$ of an alignment path $Z$ between two feature sequences can now then be measured as

$$p_{\text{align}}(Z) = \prod_{n \in Z} p_d(n) = \prod_{n \in Z} \mathcal{N}_{\text{cdf}}(d(\mathbf{x}_{p(n)}, \mathbf{x}_{q(n)}) | \mu_d, \sigma_d^2) \quad (1)$$

where $p(t)$ and $q(t)$ denote pointers from the alignment path to the original feature data $\mathbf{X}$ for the two segments-in-alignment $i$ and $j$, respectively. In short, the smaller the $p_{\text{align}}$, the more unlikely it is that an alignment of this quality would occur by chance. Note that the full alignment path across $\mathbf{P}_{i,j}$ is unlikely to correspond to a matching linguistic pattern. Instead, we assume that such a unit may lie somewhere on the path. The *quality* of any given sub-path along the full alignment path is therefore determined by measuring a likelihood ratio (LR) between $p_{\text{align}}$ and $p_{\text{chance}}$ for the sub-path. The latter is a surrogate measure defined as $p_{\text{chance}} = \alpha^{|Z|}$, where $|Z|$ denotes sub-path length. $\alpha$ is the same hyperparameter as in stage #1, now denoting the confidence level that the alignment is not caused by chance. Formally, LR can be written as:

$$LR(Z) = \frac{p_{align}(Z)}{p_{chance}(Z)} \propto \log(p_{align}(Z)) - \log(p_{chance}(Z))$$
$$= \sum_{n \in Z} \log(p_d(n)) - \sum_{t \in T} \log(\alpha) \quad (2)$$

LR is calculated separately for all possible sub-paths $Z \in \Omega_{i,j}$ on the full alignment path between the extended segments $i$ and $j$ and the path minimizing the LR is chosen as the final alignment. If the resulting alignment is shorter than $L_{min}$ steps, the pair is discarded. This also leads to automatic pruning of poor candidates from stage #1, as their LR will reach the minimum for short sub-paths.

---

[1] Large enough sample to have corvergence in the distribution parameters, here at the scale of $10^6$ data points.

[2] For DTW, we used *dp2* implementation for MATLAB by Dan Ellis, available at https://labrosa.ee.columbia.edu/matlab/dtw/.

As for computational complexity, the number of pairwise distances in stage #1 grows with $O(n^2)$ while stage #2 is $O(n)$. Both stage 1 and stage 2 calculations can be trivially parallelized, providing a linear speedup with additional CPUs.

## 3. Experiments

### 3.1. Data

Performance of the PDTW was evaluated as a part of ZSC2020. Data for this challenge comes in the form of five corpora, where three (Mandarin, French, and English) are provided to the participants together with evaluation software to test the performance of the developed systems. In addition, two surprise languages ("LANG1" and LANG2") were included. For these, evaluation was only carried out by the challenge organizers for the official challenge submission, allowing two system submissions per participant in total. Details of the datasets are available on challenge website at http://www.zerospeech.com.

### 3.2. Evaluation

Pattern matching was evaluated using the ZSC2020 evaluation toolkit (see [8]). We report normalized edit distance (NED) of the phoneme strings corresponding to the discovered pattern pairs, coverage of the discovered patterns (Cov) w.r.t. all speech in the given corpus, and word segmentation performance (precision, recall, F-score) comparing discovered patterns with ground-truth word boundaries. Since "*grouping*" measure of the evaluation toolkit—initially meant for pattern matching quality— was not computable for moderate pattern counts even with a powerful cluster, we also introduce a new total matching score to summarize the overall tradeoff between coverage and purity. More specifically, we computed the harmonic mean of the two primary measures of NED and Cov:

M-score = 2 × (100–NED) × Cov/(100–NED+Cov)     (3)

M-score reaches its maximum value of 100 if coverage is 100% and NED is 0, i.e., when all speech of the corpus has been assigned to at least one matching pair while all discovered pairs have exactly the same phonemic content. In contrast, it will obtain a value of 0 if one or both of the primary measures get the worst possible value.

### 3.3. Experimental setup

Input features to PDTW were standard mean and variance normalized 39-dim MFCCs (13 static + delta + deltadelta). We also explored the use of autoregressive predictive coding (APC; [23]) features trained on the same data. However, the very high dimensionality of APC features ($d$ = ~300 after PCA) on English and French datasets caused some computational issues for our experiment timeline. We therefore report performance on all corpora using the first 39 principal coefficients of the APC features ("APC39") to compare them with MFCCs of equal dimensionality, and only report full APC performance for the smallest Mandarin dataset ("APCfull").

As for PDTW parameters, we used window of $L$ = 20 and shift of $S$ = 10 frames (200 & 100 ms), downsampling to $M$ = 4 frames, keeping up to $k$ = 5 best matches for each segment, and expanding original segments by $E$ = ±25 frames in stage #2.

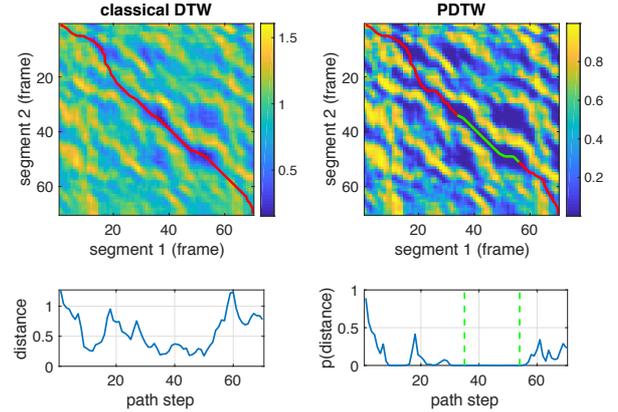

Figure 2: *Examples of classical DTW (top left) vs. probabilistic DTW (top right) affinity matrices with minimum cost paths. Corresponding path costs are on bottom left and right, respectively. Green segment denotes the best PDTW sub-path.*

Alignments shorter than 50 ms ($L_{min}$ = 5) were discarded. Experiments were conducted for two significance levels of α = 0.001 and α = 0.0001. Instead of using the official challenge oracle VAD to discard non-speech frames, we used a 2-component GMM fit to all data on $0^{th}$ MFCC coefficient (or $1^{st}$ PCA coefficient for APC) to discard all frames that had one-tailed probability of p < 0.01 of belonging to the larger cluster (i.e., the cluster assumed to contain speech frames).

For surprise languages LANG1 and LANG2, only PDTW-MFCC results are reported, as only two evaluation submissions were allowed in the challenge per participant.

### 3.4. Reference methods

As a reference, we report results for JHU-UTD system [17], which is the official baseline of the ZSC2020, and for system submission no. #1 from S. Bhati [24] that was available on the challenge leaderboard at the time of writing. In addition, we show results for ES-Kmeans [22] and systems by García and Sanchis [25] (from now on: "GS"), and Räsänen and Seshadri [20] (from now on: "Syl-DTW"), all originally submitted to ZS2017, and for which updated ZSC2020 results are also available on the challenge website. JHU-UTD and Syl-DTW are the most important comparison points, as they both use DTW in their core functionality. In contrast, the ES-Kmeans always performs a full segmentation and clustering of the corpus while GS uses supervised Hungarian acoustic model as a system component. Performance comparison to the last two should therefore be carried out with reservation.

## 4. Results

Table 1 shows the results for ZS2020 Track-2 evaluations on the ZS2017 datasets. Our primary challenge submission with α = 0.001 is highlighted with a green background, and the official challenge baseline is highlighted with yellow background.

As can be observed from the table, PDTW-MFCC (α = 0.001) has relatively consistent performance across all five languages, including the two surprise languages. Purity of the pairings is consistent and high compared to all other non-baseline systems. At the same time, the PDTW system still covers a substantial proportion of each corpus (M-score ranging from 55.3 on English to 72.0 on French). In comparison, the baseline system finds very high-quality pairings on four out of five languages but fails totally on French, and only covers a very small proportion of each corpus (M-scores ranging from 4.4 to 14.1). Word segmentation scores of PDTW are also consistent and comparable, but not superior, to those obtained by ES-Kmeans. When α is set to a very stringent value of 0.0001, PDTW finds less patterns that are more pure, as would be expected. The average NED is now lower than that of the baseline system while reaching much higher coverage on four of the five corpora.

Table 1: *Results for ZSC2020 Track-2 evaluations. NED = normalized edit distance, Cov = coverage of discovered terms w.r.t. entire corpus, M = M-score. Word segmentation measures: PRC = precision, RCL = recall, F = F-score. Our primary submission to ZSC2020 challenge is highlighted with green and official challenge baseline with yellow background.*

| | | General | | | Boundary | | |
|---|---|---|---|---|---|---|---|
| **Mandarin** | Words | Ned | Cov | M | PRC | RCL | F |
| PDTW-MFCC α = 0.001 | 9630 | 57.6 | 79.6 | **55.3** | 34.2 | 87.4 | 49.1 |
| PDTW-MFCC α = 0.0001 | 709 | 34.9 | 10.4 | 17.9 | 34.7 | 15.5 | 21.4 |
| PDTW-APC39 α = 0.001 | N/A | 67.1 | 69.2 | 44.6 | 32.1 | 76.5 | 45.2 |
| PDTW-APC39 α = 0.0001 | N/A | N/A | 0.0 | N/A | N/A | N/A | N/A |
| PDTW-APCfull α = 0.001 | N/A | 69.8 | 86.2 | 44.7 | 32.7 | 87.8 | 47.7 |
| García & Sanchis #1 (2017) | 2887 | 80.3 | 42.0 | 26.8 | 38.3 | 29.1 | 33.1 |
| Räsänen & Seshadri (2017) | 26529 | 67.7 | 35.4 | 33.8 | 29.7 | 36.3 | 32.7 |
| Kamper et al. (2017) | 2967 | 82.0 | 100.0 | 30.5 | 42.6 | 75.6 | **54.5** |
| Bhati #1 (2020) | 37457 | 94.7 | 99.9 | 10.1 | 36.5 | 91.9 | 52.2 |
| Jansen & Van Durme (2011) | 267 | 28.6 | 2.7 | 5.2 | 54.3 | 1.3 | 2.5 |
| **French** | Words | Ned | Cov | M | PRC | RCL | F |
| PDTW-MFCC α = 0.001 | 42988 | 36.7 | 83.5 | **72.0** | 27.9 | 88.7 | 42.5 |
| PDTW-MFCC α = 0.0001 | 2604 | 20.3 | 17.5 | 28.7 | 27.9 | 25.1 | 26.5 |
| PDTW-APC39 α = 0.001 | N/A | 58.8 | 99.6 | 58.3 | 27.9 | 97.8 | **43.5** |
| PDTW-APC39 α = 0.0001 | N/A | 53.8 | 77.3 | 57.8 | 27.5 | 83.3 | 41.3 |
| García & Sanchis #1 (2017) | 58701 | 64.3 | 77.7 | 48.9 | 29.4 | 49.7 | 36.9 |
| Räsänen & Seshadri (2017) | 195959 | 46.2 | 30.7 | 39.1 | 27.5 | 31.8 | 29.5 |
| Kamper et al. (2017) | 28733 | 66.7 | 100.0 | 50.0 | 26.9 | 44.0 | 33.4 |
| Bhati #1 (2020) | 135048 | 89.0 | 99.8 | 19.8 | 28.3 | 81.1 | 41.9 |
| Jansen & Van Durme (2011) | 1963 | 69.5 | 1.6 | 3.0 | 32.5 | 0.6 | 1.2 |
| **English** | Words | Ned | Cov | M | PRC | RCL | F |
| PDTW-MFCC α = 0.001 | 85425 | 48.2 | 85.4 | **64.5** | 26.5 | 88.2 | 40.8 |
| PDTW-MFCC α = 0.0001 | 6308 | 30.4 | 23.1 | 34.7 | 26.0 | 31.1 | 28.3 |
| PDTW-APC39 α = 0.001 | N/A | 64.8 | 99.8 | 52.0 | 26.3 | 96.6 | 41.3 |
| PDTW-APC39 α = 0.0001 | N/A | 58.0 | 88.4 | 56.9 | 25.7 | 88.5 | 39.9 |
| García & Sanchis #1 (2017) | 92544 | 72.3 | 76.8 | 40.7 | 27.8 | 45.5 | 34.5 |
| Räsänen & Seshadri (2017) | 321603 | 52.5 | 28.7 | 35.8 | 25.8 | 29.9 | 27.7 |
| Kamper et al. (2017) | 42473 | 72.3 | 100.0 | 43.4 | 39.6 | 61.4 | **48.2** |
| Bhati #1 (2020) | 240033 | 89.5 | 99.5 | 19.0 | 27.3 | 75.9 | 40.1 |
| Jansen & Van Durme (2011) | 18821 | 32.4 | 7.9 | 14.1 | 32.1 | 3.2 | 5.9 |
| **LANG1** | Words | Ned | Cov | M | PRC | RCL | F |
| PDTW-MFCC α = 0.001 | 49268 | 40.9 | 81.9 | **68.7** | 22.5 | 88.3 | 35.9 |
| PDTW-MFCC α = 0.0001 | 3685 | 23.6 | 15.4 | 25.6 | 22.2 | 21.8 | 22.0 |
| García & Sanchis #1 (2017) | 60648 | 62.5 | 78.6 | 50.8 | 21.9 | 43.8 | 29.2 |
| Räsänen & Seshadri (2017) | 223188 | 57.7 | 31.1 | 35.8 | 21.5 | 32.2 | 25.8 |
| Kamper et al. (2017) | 28675 | 68.9 | 99.9 | 47.4 | 32.5 | 57.5 | **41.5** |
| Bhati #1 (2020) | 158045 | 89.6 | 99.8 | 18.8 | 23.4 | 80.7 | 36.2 |
| Jansen & Van Durme (2011) | 4370 | 32.1 | 3.2 | 6.1 | 28.8 | 1.3 | 2.5 |
| **LANG2** | Words | Ned | Cov | M | PRC | RCL | F |
| PDTW-MFCC α = 0.001 | 1491 | 28.8 | 38.3 | **49.8** | 31.3 | 49.5 | 38.4 |
| PDTW-MFCC α = 0.0001 | 58 | 10.1 | 0.5 | 1.0 | 33.9 | 0.8 | 1.6 |
| García & Sanchis #1 (2017) | 5468 | 57.1 | 56.3 | 48.7 | 37.9 | 37.4 | 37.7 |
| Räsänen & Seshadri (2017) | 2103 | 26.1 | 20.0 | 31.5 | 33.0 | 21.9 | 26.3 |
| Kamper et al. (2017) | 12599 | 72.4 | 99.9 | 43.3 | 47.9 | 59.0 | **52.6** |
| Bhati #1 (2020) | 22221 | 89.2 | 64.1 | 18.5 | 24.8 | 55.8 | 34.4 |
| Jansen & Van Durme (2011) | 575 | 29.6 | 3.4 | 6.5 | 45.8 | 1.5 | 2.9 |

Fig. 3 illustrates the distributions of pattern lengths for the five tested languages with α = 0.001. As can be seen, the patterns are of similar length in all languages, where a substantial proportion of them are 100–300 ms, i.e., typical syllable (or monosyllabic word) lengths. In contrast, the baseline system only finds relatively long fragments of speech. In principle, this is a desirable property, as high precision of word boundaries in the baseline shows that it is focusing on long word-like units. On the other hand, it misses a large proportion of short words and other phonemic patterns. On French, PDTW finds more long patterns than the baseline system but with a substantially lower overall NED (Table 1).

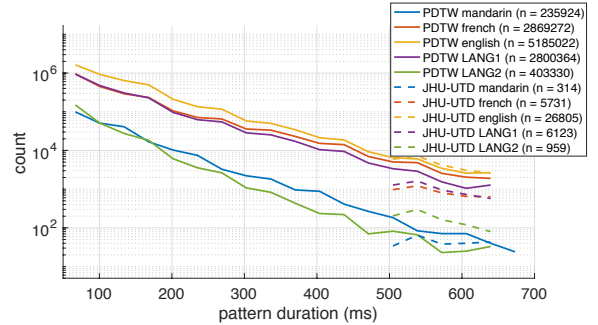

Figure 3: *Length distributions of the discovered patterns in different languages for PDTW and JHU-UTD. Total pattern counts in each language are also shown in the legend.*

Interestingly, we found no systematic improvement in performance when using APC-based features, even though the same features clearly outperform MFCCs on phonemic selectivity on Track-1 evaluations of the challenge [26]. The reason for this is currently unclear, and would require further investigation.

## 5. Conclusions

This paper described a probabilistic variant for DTW-based unsupervised spoken term discovery. The results from ZS2020 challenge evaluation show that the PDTW algorithm has consistent performance across five different languages using the same hyperparameters. It also clearly outperforms the official challenge baseline system also using DTW for pattern matching, especially when balance between coverage and purity of the discovered pairings is considered. PDTW also outperforms the earlier Syl-DTW system [20] that uses a similar pipeline but with heuristic thresholding and syllabic instead of fixed-length input segments. In addition, PDTW finds more word-like patterns than the two reference methods, as reflected by the higher boundary F-score on the two test languages. Overall, the results show that the adaptive behavior obtained through probabilistic modeling of feature distances and DTW-alignment path cost estimation resolves the issue of proper hyperparameter tuning for UTD in different corpora—a major concern in the earlier approaches. In principle, similar formulation should apply to other multivariate time-series pattern discovery tasks also outside speech domain.

In the future work, PDTW system could be expanded to incorporate proper mechanisms for grouping of the discovered pairs into clusters of acoustic patterns. It could also be used as a front-end in systems such as correspondence autoencoders [27] for low-resource feature learning or as a potential replacement for frame alignment in INCA algorithm [28] in voice and speaking style conversion systems trained on non-parallel data, i.e., whenever it is beneficial to find alignments of longer recurring patterns in a corpus (or corpora) of speech.

## 6. Acknowledgements

This research was funded by Academy of Finland grants no. 314602 and 320053. Code for the PDTW is available for download at https://github.com/SPEECHCOG/ZS2020

# 7. References

[1] H. Kamper, A. Jansen, & S. Goldwater. Unsupervised word segmentation and lexicon discovery using acoustic word embeddings. *IEEE/ACM Trans. Audio, Speech, and Language*, vol. 24, no. 4, pp. 669–679, 2016.

[2] H. Kamper, A. Jansen, & S. Goldwater. A segmental framework for fully-unsupervised large-vocabulary speech recognition. *Computer Speech & Language*, vol. 46, pp. 154–174, 2017.

[3] E. Dupoux. Cognitive science in the era of artificial intelligence: A roadmap for reverse-engineering the infant language learner. *Cognition*, vol. 173, pp. 43–59, 2018.

[4] O. Räsänen. Computational modelling of phonetic and lexical learning in early language acquisition: Existing models and future directions. *Speech Communication*, vol. 54, no. 9, pp. 975–997, 2012.

[5] M. Versteegh, R. Thiollière, T. Schatz, X.N. Cao, X. Anguera, A. Jansen, & E. Dupoux. The zero resource speech challenge 2015. *Proc. Interspeech-2015*, September 6–10, Dresden, Germany, 2015, pp. 3169–3173.

[6] E. Dunbar, X-N. Cao, J. Benjumea, J. Karadayi, M. Bernard, L. Besacier, X. Anguera, & E. Dupoux. The zero resource speech challenge 2017. *Proc. IEEE Workshop on Automatic Speech Recognition and Understanding (ASRU-2017)*, Okinawa, Japan, 2017, pp. 323–330.

[7] E. Dunbar et al.. The zero resource speech challenge 2019: TTS without T. *Proc. Interspeech-2019*, Sep. 15–19, Graz, Austria, 2019, pp. 1088–1092.

[8] http://www.zerospeech.com/2020/news.html. <to be replaced with challenge paper when available>

[9] H. Sakoe & S. Chiba. Dynamic programming algorithm optimization for spoken word recognition. *IEEE Trans. Acoustics, Speech, and Signal Processing*, vol. 26, no. 1, pp. 43–49, 1978.

[10] J. Ziv & A. Lempel. Compression of individual sequences via variable-rate coding. *IEEE Trans. Information Theory*, vol. 24, pp. 530–536, 1978.

[11] J. Rissanen. Modeing by shortest data description. *Automatica*, vol. 14, no. 5, pp. 465–658, 1978.

[12] K. Jensen, M. Styczynski, I. Rigoutsos, & G. Stephanopoulos. A generic motif discovery algorithm for sequential data. *Bioinformatics*, vol. 22, no. 1, pp. 21–28, 2006.

[13] O. Räsänen. A computational model of word segmentation from continuous speech using transitional probabilities of atomic acoustic events. *Cognition*, vol. 120, no. 2, pp. 149–176, 2011.

[14] A. Park & J. Glass. Unsupervised word acquisition from speech using pattern discovery. *Proc. ICASSP-2006*, May 14–19, Toulouse, France, 2006, pp. 409–412.

[15] L. ten Bosch & B. Cranen. A computational model for unsupervised word discovery. *Proc. Interspeech*-2007, Antwerp, Belgium, 2007, pp. 1481–1484.

[16] G. Aimetti. Modelling early language acquisition skills: Towards a general statistical learning mechanism. *Proc. SRW-EACL-2009*, April 2, Athens, Greece, 2009, pp. 1–9.

[17] A. Jansen & B. Van Durme. Efficient spoken term discovery using randomized algorithms. *Proc. IEEE Workshop on Automatic Speech Recognition & Understanding (ASRU-2011)*, December 11–15, Hawaii, USA, 2011, pp. 401–406.

[18] L. Cantanese, N. Souviraà-Labastie, B. Qu, S. Campion, G. Gravier, E. Vincent & F. Bimbot. MODIS: an audio motif discovery software. *Proc. Interspeech-2013*, August 25–29, Lyon, France, 2013, pp. 2675–2677.

[19] A. Muscariello, G. Gravier, & F. Bimbot. Unsupervised motif acquisition in speech via seeded discovery and template matching combination. *IEEE Trans. Audio, Speech, and Language Procssing*, vol. 20, no. 7, September, pp. 2031–2044, 2012.

[20] O. Räsänen & S. Seshadri. ZS2017 AaltoLAG Submission #1 [Data set]. Zenodo. http://doi.org/10.5281/zenodo.810808, 2017.

[21] O. Räsänen, G. Doyle & M.C. Frank. Unsupervised word discovery from speech using automatic segmentation into syllable-like units. *Proc. Interspeech-2015*, Dresden, Germany, 2015, pp. 3204–3208.

[22] H. Kamper, K. Livescu, & S. Goldwater. An embedded segmental k-means model for unsupervised segmentation and clustering of speech. *Proc. IEEE Workshop on Automatic Speech Recognition & Understanding (ASRU-2017)*, December 16–20, Okinawa, Japan, 2017, pp. 719–726.

[23] Y-A. Chung, W-N. Hsu, H. Tang, & J. Glass. An unsupervised autoregressive model for speech representation learning. *Proc. Interspeech-2019*, September 15–19, Graz, Austria, 2019, pp. 146–150.

[24] S. Bhati. Self clustering autoencoder unsupervised features learning. System submission #1 for Zero Resource Speech Challenge 2020 held at *Interspeech-2020*, October 25–29, Shanghai, China, 2020.

[25] F. García & E. Sanchis. ZS2017 ELIRF Submission #1 [Data set]. Zenodo. http://doi.org/10.5281/zenodo.815468, 2017.

[26] M. A. Cruz Blandón, & O. Räsänen. Analysis of predictive coding models for phonemic representation learning in small datasets. *Proc. ICML 2020 Workshop on Self-supervision in Audio and Speech*, held as a virtual conference, 2020.

[27] P-J. Last, A. Engelbrecht, & H. Kamper. Unsupervised feature learning for speech using correspondence and Siamese networks. *IEEE Signal Processing Letters*, vol. 27, pp. 421–425, 2020.

[28] D. Erro, A. Moreno, & A. Bonafonte. INCA algorithm for training voice conversion systems from nonparallel corpora. *IEEE Trans. Audio, Speech, and Language Processing*, vol. 18, no. 5, pp. 944–953, 2010.